# The Absurdity of Death Estimates Based on the Vaccine Adverse Event Reporting System

Gordon V. Cormack, Ph.D.
Maura R. Grossman, J.D., Ph.D.

*David R. Cheriton School of Computer Science*
*University of Waterloo*

**KEY MESSAGES**

- **Preposterous death estimates from the Vaccine Adverse Event Reporting System ("VAERS") data are logically and mathematically unsound**
- **VAERS reporting rate cannot be used to infer incidence of death or any other adverse event following vaccination**
- **VAERS statistics do not show increased deaths after COVID vaccination**
- **VAERS statistics do not show any COVID vaccine-caused deaths after vaccination**

**Abstract**

*We demonstrate from first principles a core fallacy employed by a coterie of authors who claim that data from the Vaccine Adverse Reporting System ("VAERS") show that hundreds of thousands of U.S. deaths are attributable to COVID vaccination.*

**Introduction**

Arguments [1-4] derived from VAERS [5] that COVID vaccines have killed vast numbers of people fail for one principal reason: They hinge on the false assumption that the reporting rate for adverse events following vaccination is strictly proportional to the actual incidence of such events. Rose [1] introduces the term "underreporting factor," $URF = \frac{incidence}{reporting\ rate}$, which she claims to be invariant for all serious adverse events ("SAEs"). Based on an estimate of $URF = 31$, she claims to show: "205,809 dead, 818,462 hospitalizations, 1,830,891 ER visits, 230,113 life-threatening events, 212,691 disabled and 7,998 birth defects to date." Several authors have advanced this and similar arguments [1-4], all relying either tacitly or explicitly on the false assumption that URF is constant notwithstanding differences in the type of vaccine, the nature and cause of the event, or the time elapsed from vaccination to event.

These authors reject the prominent disclaimer [5] that must be acknowledged to access VAERS data: "The number of reports alone cannot be interpreted or used to reach conclusions about the existence, severity, frequency, or rates of problems associated with vaccines." A number of non-archival works [6-9] have rebutted these false claims, citing authorities, evidence to the contrary, and sophisticated data analysis. Members of the coterie have in turn dismissed these rebuttals as the product of conspiracy or the work of ivory-tower academics disconnected from reality (See Box 1). The purpose of this article is to show, from first principles, using only basic mathematics and elementary logic that can be understood without specialized training, that the foundational assumption on which the false claims depend is not supported by the data. This analysis may be useful as a tool to help convince some of those who demand "first-hand" evidence that that COVID vaccine is not "a biological product that has killed over 150,000 Americans to date" [3].

---

**Box 1: Doubling Down on Disinformation: Some Examples**

Kostoff [10]: *"Given the blatant censorship of the mainstream media and social media, only one side of the COVID-19/"vaccine" narrative is reaching the public. Any questioning of the narrative is met with the harshest response. Front-line people (doctors, nurses, etc.) who are attempting to shed light on this situation are being fired, losing licenses, and having their reputations and finances destroyed."*

Kirsch, Crawford & Rose [3]: *"It's all driven by the CDC false narrative that there are no deaths. Nobody questions that. Nobody until now has even asked the FDA or CDC for the URF . . . I was the first one. Are you kidding me? Nobody in the medical community wants to know the URF?!? So yeah, that's proof that people have their head in the sand and aren't interested in safety signals."*

Kirsch, Crawford & Rose [3]: *"The difference between Professor Morris and me is that I will take what is on the table, and spend money to do things like professional polls and physician surveys to gather more data and then do my best to analyze it."*

Crawford [11]: *"It's baffling that a computer scientist with real computations is this insistent on cluelessness."*

---

**The Vaccine Adverse Event Reporting System**

VAERS is a passive system that captures reports of deaths or other adverse events following vaccination. For vaccines other than COVID vaccines, healthcare providers are required by law [12] only to report prescribed events within a specific number of days of vaccination, as well as "any acute complication or sequelae (including death) of above events." For COVID vaccines, however, providers are required by law [13] to report all "Serious AEs (including death) *regardless of causality*" after vaccination, with no specified time interval for what constitutes "after vaccination." (Emphasis added.)

**Data Sources**

The present analysis relies on five sources of data: The VAERS dataset for vaccinations and events as of January 30, 2022 [1], daily reports of COVID vaccination by age category in the United States [14], weekly reports of all-cause and COVID-related deaths by age category in the United States [15], estimates of influenza vaccine coverage by age category in the United States [16], and estimates of zoster vaccine coverage by age category in the United States [17]. For brevity, we focus on vaccinations, deaths, and VAERS reports for the 2021 calendar year, with respect to the U.S. population aged 65 and older.

From the death and vaccination records, we know that 2.4M of 59M Americans in the 65+ age category died in 2021, an average of 6,600 deaths per day. In other words, an average of 112 per million people died per day, regardless of whether or not, or when, they had been vaccinated.

From the COVID-vaccination data, we can deduce that in 2021, 55M people in the 65+ age category received a first dose, and 48M became "fully vaccinated." Since all but a few percent of the vaccine doses administered in the U.S. were part of a two-dose regimen, we know that the total number of doses administered in 2021 was nearly the sum of these figures: roughly 103M.

Other than COVID, the two most-reported vaccines for the 65+ age category are influenza and zoster (shingles). In the 2020/21 influenza season, 57.8% of this population were vaccinated. Based on the assumption that uptake for the 2021 calendar year was similar, we estimate that 34M influenza vaccines were administered in 2021.

Survey results show that about 35% of the 65+ population is vaccinated for zoster. Since the regimen involves two doses, and lasts for five years, we assume that about one in five seniors receive the two-dose regimen per annum. Under this assumption, approximately 8.3M doses of zoster vaccine were administered in 2021.

For the 65+ age category, the VAERS dataset contains 6,604 reports of death following COVID vaccination in 2021; 29 reports of death following influenza vaccination; and 4 reports of death following zoster vaccination.

It does not require sophisticated mathematics to see that the reporting rate for deaths following vaccination is orders of magnitude greater for COVID than for influenza or zoster vaccines, while the actual incidence is comparable. A more precise analysis follows.

**Reporting Rate vs. Incidence vs. Causality**

"Reporting rate" is the number of reports of a prescribed event per person per unit time, expressed in this article as reports/million/day.  "Incidence" is the number of actual occurrences of the prescribed event per person per unit time, expressed as events/million/day.  It follows that, for a given prescribed event and interval of time, $URF = \frac{incidence}{reporting\ rate}$.

Causality is an important factor in determining the nature of an event.  In particular, the events "death following vaccination" and "death caused by vaccination" are materially different in nature, with different incidence and reporting rates.  In her claim of "205,809 dead," Rose [1] tacitly conflates such all-cause and vaccine-caused deaths.  It is vacuously true that more than 202,809 Americans have died from all causes since being vaccinated for COVID.  It does not follow that 202,809—or any—of those deaths were *caused by* vaccination.

Kostoff et al. [2] suggest that causality could be established if we knew the proportion of reported all-cause events that were in fact vaccine-caused:

> Suppose, for example, that the actual number of deaths reported in VAERS came from two groups:  90% were from the inoculation-attributable death group and 10% were from the expected death group.  Assume there is no overlap between the two groups.  In that case, what VAERS shows is not that 1% of actual expected deaths were reported, but rather that 1/10 of one percent of the expected deaths were reported.  If that metric is used as the standard to scale up to total deaths, then the number in the actual inoculation-attributable death group is not 100 times the VAERS reported deaths, but rather 1000 times the VAERS-reported deaths!

This argument is not only mathematically flawed, but precisely backwards.  If 90% of reported deaths were vaccine-caused, it would indicate that the reporting rate for vaccine-caused deaths was much greater—and the URF much smaller—than for all-cause deaths.  How much smaller is impossible to determine without knowing the actual incidence of vaccine-caused deaths.

Kirsch et al. [3], building on the same flawed reasoning, purport to show clinical evidence that a large proportion of reported events were vaccine-caused.  Even if this clinical evidence were compelling, it would show only that the reporting rate for vaccine-caused deaths was higher than for all-cause deaths.

Kirsch et al. [3] argue that causality can be established by showing "excess deaths," which they define to be the total number of deaths following COVID vaccination minus the total number of deaths following "the typical vaccine," independent of causation. In estimating the total number of deaths, they rely on the key false assumption that the URF for all-cause deaths is the same for COVID as for "the typical vaccine."

Kostoff et al. [2], referring to a plot of reported all-cause deaths as a function of days since vaccination (*cf.* Figure 1, below), argue: "If there were no effect from the inoculation, as claimed by the CDC and other official government agencies, the curve would be essentially a straight horizontal line, reflecting normal expected deaths in a non-COVID-19 year." This argument fails on two fronts: (1) it relies on the false assumption that URF for all-cause deaths is independent of the time elapsed since vaccination; and (2) the same temporal association—not a straight horizontal line as claimed—is seen for other vaccines, in 2021, and in prior years.

**Analysis**

To calculate reporting rate and incidence, we must specify a time interval. Consider first, the number of reported and actual deaths on the day immediately following vaccination. For the 2021 calendar year, among people at least 65 years of age, there are 409 reports of fatalities the day following COVID vaccination, 7 following influenza vaccination, and 1 following zoster vaccination. These numbers show reporting rates of 4/million/day, 0.2/million/day, and 0.12/million/day, respectively—all far lower than the incidence of 112 deaths/million/day in the general population. VAERS reports provide no evidence of "excess deaths," either relative to incidence in the general population, or relative to other vaccines. Rather, they indicate that all-cause deaths are underreported by a larger factor for other vaccines than for COVID vaccines.

Figure 1 shows the decline in reporting rate over the 60 days following all COVID, influenza, and zoster vaccinations from January 1, 2000 to October 31, 2021. We see that there is no material difference in falloff between the COVID vaccine and the vaccines administered in "non-COVID years." The falloff indicates that all-cause deaths are underreported by a larger factor when temporally distant from vaccination.

**Figure 1: Reporting Rate Falloff as a Function of Days Elapsed Since Vaccination**

Similar falloff for COVID and non-COVID vaccines

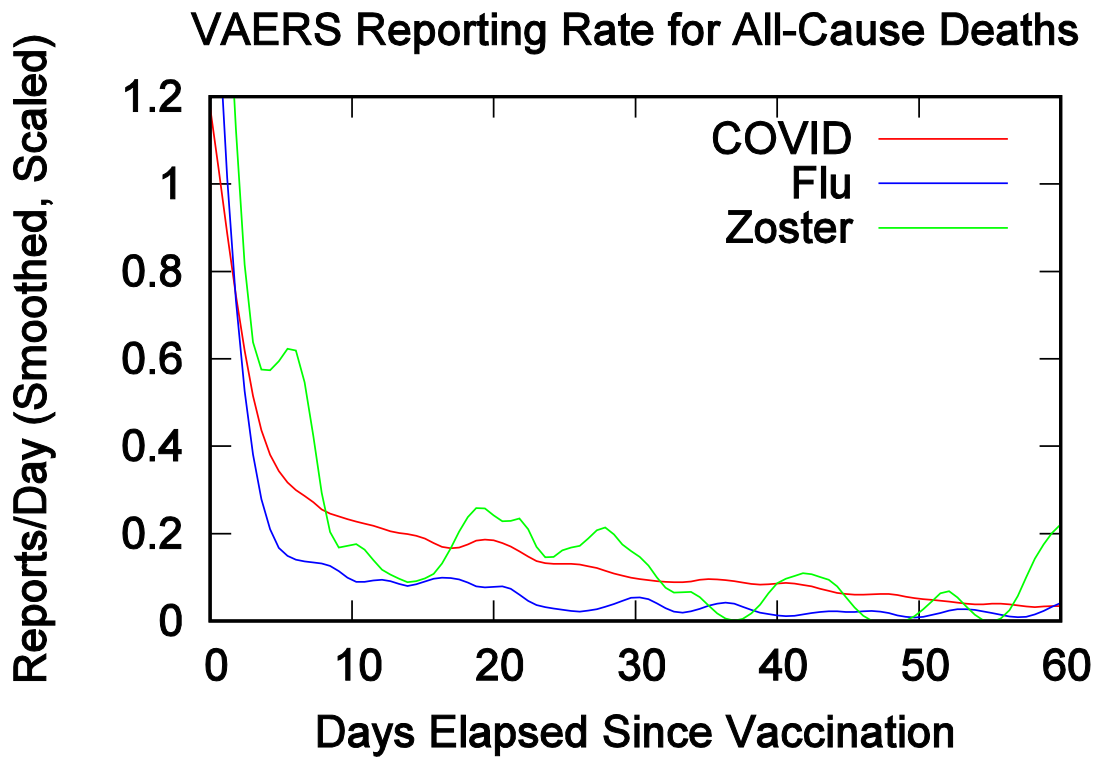

**Conclusion**

By examining reported all-cause deaths vs. known all-cause deaths, we have shown that there is no single URF by which one can multiply the number of VAERS reports to estimate the number of actual events, let alone the number of events caused by COVID vaccination. All of the arguments purporting to quantify COVID vaccine-related deaths from VAERS, and all of the arguments purporting to show causality, fail for the same reason—because they hinge on this false assumption regarding URF. Arguments relying on URF similarly fail to show evidence of elevated adverse events other than death, and hence offer no basis to disregard the clear VAERS disclaimer: "The number of reports alone cannot be interpreted or used to reach conclusions about the existence, severity, frequency, or rates of problems associated with vaccines."